\documentclass[12pt]{article}
\def\theequation{\arabic{section}.\arabic{equation}}
\usepackage{amssymb}

\newcounter{rown}
\def\bl{\setcounter{rown}{\value{equation}}
        \stepcounter{rown}\setcounter{equation}0
        \def\theequation{\thesection.\arabic{rown}\alph{equation}}
        }
\def\el{\setcounter{equation}{\value{rown}}
        \def\theequation{\thesection.\arabic{equation}}}
\def\sec{\setcounter{equation}0}
\begin{document}
\renewcommand{\thefootnote}{\fnsymbol{footnote}}
\renewcommand{\theequation}{\thesection.\arabic{equation}}
\title{The Symmetry Algebras of Euclidean $M$-theory}
\author{Jerzy Lukierski${}^{a,b}$\thanks{{\em e-mail: lukier@ift.uni.wroc.pl}}
 and Francesco Toppan${}^b$\thanks{{\em e-mail: toppan@cbpf.br}}
\\ \\
${}^a${\it Institute for Theoretical Physics, University of
Wroc{\l}aw,}
\\ {\it 50-204 Wroc{\l}aw, pl. Maxa Borna 9, Poland}\\
 \\ ${}^b${\it CBPF, CCP, Rua Dr.}
{\it Xavier Sigaud 150,}
 \\ {\it cep 22290-180 Rio de Janeiro (RJ), Brazil}
}

\maketitle
\begin{abstract}
We study the Euclidean supersymmetric $D=11$ $M$-algebras.
 We consider two such
   $D=11$ superalgebras: the
   first one is  $N=(1,1)$ self-conjugate complex-Hermitean,
   with $32$
complex supercharges and $1024$ real bosonic charges,
  the second is $N=(1,0)$ complex-holomorphic, with
 $32$ complex supercharges and $528$ bosonic charges,
  which can be obtained
  by analytic continuation of known Minkowski  $M$-algebra.
    Due to the
Bott's periodicity, we study at first the generic $D=3$ Euclidean
supersymmetry case. The role of complex and quaternionic
structures for $D=3$ and  $D=11$ Euclidean supersymmetry is
elucidated. We show that the additional $1024-528=496$ Euclidean
tensorial
 central charges are related with the quaternionic structure of
 Euclidean  $D=11$ supercharges, which in complex notation  satisfy
  $SU(2)$ pseudo-Majorana condition. We consider also the
 corresponding Osterwalder-Schrader conjugations
   as implying for $N=(1,0)$ case the reality of
     Euclidean bosonic charges.
  Finally, we outline some consequences of our results, in
 particular for $D=11$ Euclidean supergravity.
\end{abstract}

\newpage

\section{Introduction.}

The physical spacetime is Minkowskian, but there are several reasons
 justifying the interest in Euclidean theories. We can
  recall here that

{\em i}) The functional integrals acquire a precise mathematical
meaning only in the context of Euclidean quantum
 theory   (see e.g. \cite{ref1,ref2}).
  \par
{\em ii}) The  topologically nontrivial field configurations
 (such as instantons)
are solutions of Euclidean field theories (see e.g.
\cite{ref3,ref4}).
\par
{\em iii})  The generating functional of
 an Euclidean field theory can be related
to the description of statistical and stochastic systems
(\cite{ref5,ref6}).

At present the $D=11$ $M$-theory has been considered as the most recent
  proposal for a ``Theory of
Everything'' ( see e.g. \cite{ref7,ref8}). We still do not know
the dynamical content
 of the $M$-theory,
however it seems that the algebraic description of its symmetries
is well embraced by the so-called $M$-algebra\footnote{In fact it
is a superalgebra, also sometimes called $M$-superalgebra.}
\cite{ref9,ref10}
\begin{eqnarray}\label{lutop1}
    \left\{ Q_A, Q_B \right\} & = & P_{AB}
     = \left( C \Gamma_\mu \right)_{AB} P^\mu
     \cr
     && + \ \left( C \Gamma_{[\mu\nu]}\right)_{AB}
     Z^{[\mu\nu]} +
     \ \left( C \Gamma_{[\mu_1 \ldots \mu_5]}\right)_{AB}
     Z^{[\mu_1 \ldots \mu_5]}
\end{eqnarray}
where the $Q_A$ ($A=1,2,\ldots , 32$) are
 $32$ $D=11$ real Minkowskian supercharges, $P_\mu$ describe the
$11$-momenta, while the remaining $517$ bosonic
 generators describe the tensorial central charges
  $Z^{[\mu\nu]}$ and $Z^{[\mu_1 \ldots \mu_5]}$.

It should be stressed that the $M$-algebra (\ref{lutop1}) is the
generalized $D=11$ Poincar\'{e} algebra
   with maximal number of additional
bosonic generators. These additional generators indicate the
presence of $D=11$
 $M2$ and $M5$ branes. Indeed, it has been shown
   (see e.g. \cite{ref11,ref12}) that
   $D=11$  supergravity contains the super-$2$-brane (supermembrane) and
 super-$5$-brane solutions.

 Our  aim here is to study the Euclidean counterpart of the $M$-algebra,
  described by the relation (\ref{lutop1}). The problem of Euclidean
  continuation of superalgebras is not trivial, because the
  dimensionality of Minkowski and Euclidean spinors may differ, as
  it is well-known from $D=4$ case (see e.g. \cite{ref13}--\cite{ref16}).
   In a four-dimensional world
 the Minkowski spinors are ${\bf C}^2$
  (two-dimensional Weyl spinors) which
  can be described as ${\bf R}^4$ Majorana spinors, but the
fundamental $D=4$ Euclidean spinors are described by ${\bf H}
\otimes {\bf H} ={\bf H}^2$
  (due to the relation $O(4) =O(3)\times O(3)$ we deal with a pair of
  $D=3$ Euclidean spinors)
or ${\bf C}^2 \otimes {\bf C}^2= {\bf C}^4$, i.e. the number of
spinor components is doubled. Further, one can describe the
 $D=4$ Minkowski Dirac matrices as real four-dimensional
  ones (the so-called  Majorana representation), but the
  four-dimensional $D=4$ Euclidean Dirac
 matrices are necessarily complex. The doubling of spinor
 components is reflected in the analytic continuation procedure, and the
 reality condition in $D=4$ Minkowski space is replaced
   by the so-called Osterwalder-Schrader reality condition
     \cite{ref16,ref17}).

 In  $D=11$
Minkowski case  the fundamental spinors are
 ${\bf R}^{32}$, and the corresponding fundamental representation
of the Minkowskian $D=11$ Clifford algebra

\begin{equation}\label{lutop2}
    \left\{ \Gamma_\mu , \Gamma_\nu \right\} = 2\eta_{\mu \nu}\, ,
     \qquad \eta_{\mu\nu} = (-1, 1, \ldots , 1) \, ,
\end{equation}
is ${\bf R}^{32}\times {\bf R}^{32}$, which allows writing
  the $M$-algebra (\ref{lutop1}) as a real algebra.
In the $D=11$ Euclidean case the fundamental spinors
 are ${\bf H}^{16}$, while the fundamental
 Hermitean Clifford algebra representation in $D=11$ Euclidean
 space
  is ${\bf C}^{32}\times {\bf C}^{32}$.
   We shall  describe the Euclidean $M$-algebra
 using Hermitean products of   complex Hermitean $D=11$ Euclidean
 gamma-matrices satisfying the Euclidean counterpart
  of algebraic relations (\ref{lutop2}):

\begin{eqnarray}
\label{lutop3}
    \left\{ \widetilde{\Gamma}_\mu , \widetilde{\Gamma}_\nu \right\}
    = 2\, \delta_{\mu \nu}\, ,
     \qquad {\mu,\nu} = (1, \ldots , 11) \,
\end{eqnarray}
If we introduce 32 complex supercharges $\widetilde{Q}_A$
we get the following formula for the Euclidean $D=11$ superalgebra
\begin{eqnarray}\label{lutop4}
\left\{ \widetilde{Q}_A, \widetilde{Q}_B^+ \right\}
& = &
\delta_{AB} \widetilde{Z} +
\left( \widetilde{\Gamma}_\mu \right)_{AB} \widetilde{P}_\mu
+
\left( \widetilde{\Gamma}_{[\mu_1 \ldots \mu_4]}\right)_{AB}
\widetilde{Z}_{[\mu_1 \ldots \mu_4]}
\cr
&& + \
\left( \widetilde{\Gamma}_{[\mu_1 \ldots \mu_5]}\right)_{AB}
\widetilde{Z}_{[\mu_1 \ldots \mu_5]}
+
\left( \widetilde{\Gamma}_{[\mu_1 \ldots \mu_8]}\right)_{AB}
\widetilde{Z}_{[\mu_1 \ldots \mu_8]}
\cr
&&
+ \
\left( \widetilde{\Gamma}_{[\mu_1 \ldots \mu_9]}\right)_{AB}
\widetilde{Z}_{[\mu_1 \ldots \mu_9]}
\end{eqnarray}
where  on r.h.s. of (\ref{lutop4})  all linearly
independent Hermitean antisymmetric products of
 $\Gamma$-matrices appear.

Since in $D=11$ Euclidean space we get the relation
\begin{equation}\label{lutop1.5}
    \widetilde{\Gamma}_{12} = \frac{1}{11!}\, \epsilon_{\mu_1 \ldots
    \mu_{11}}\, \widetilde{\Gamma}_{\mu_1}\ldots \widetilde{\Gamma}_{\mu_{11}} = i
\end{equation}
we obtain the identity
\begin{equation}\label{lutop1.6}
    \widetilde{\Gamma}_{[\mu_{1} \ldots \mu_{k}]}
    = \frac{i}{(11-k)!}\, \epsilon_{\mu_{1} \ldots
    \mu_{11}}\,
    \widetilde{\Gamma}_{[\mu_{k+1}\ldots \mu_{11}]}\, .
\end{equation}
Applying (\ref{lutop1.6}) for $k=2$ and $3$ one can write the relation
 (\ref{lutop4}) as follows
 \begin{eqnarray}\label{lutop1.7}
   \left\{ \widetilde{Q}_A, \widetilde{Q}^{+}_{B}\right\} &
   =&  \delta_{AB} \widetilde{Z} + (\widetilde{\Gamma}_\mu)_{AB} \widetilde{P}_\mu
+ (\widetilde{\Gamma}_{[\mu\nu]})_{AB} \widetilde{Z}_{[\mu\nu]}
    +   (\widetilde{\Gamma}_{[\mu\nu\rho]})_{AB} \widetilde{Z}_{[\mu\nu\rho]}
    \cr
   && +
    (\widetilde{\Gamma}_{[\mu_1 \ldots \mu_4]})_{AB}
     \widetilde{Z}_{[\mu_1 \ldots \mu_4]}
   +  (\widetilde{\Gamma}_{[\mu_1 \ldots \mu_5]})_{AB}
   \widetilde{Z}_{[\mu_1 \ldots \mu_5]}\, ,
 \end{eqnarray}
 where
 \begin{eqnarray}\label{lutop1.8}
   \widetilde{Z}_{[\mu\nu]} &=&
\frac{i}{2!} \epsilon_{\mu\nu \nu_1 \ldots \nu_9}
\widetilde{Z}_{[\nu_1 \ldots \nu_9]}
    \cr
\widetilde{Z}_{[\mu\nu\rho]} &=&
\frac{i}{3!} \epsilon_{\mu\nu\rho \nu_1 \ldots \nu_8}
\widetilde{Z}_{[\nu_1 \ldots \nu_8]}\, .
 \end{eqnarray}

We see that in (\ref{lutop1.7}) we have two sets of Abelian
bosonic charges,
also called tensorial central charges
\par
{\em i}) The  $528=11+ 55+462$ charges
 $\widetilde{P}^\mu$, $\widetilde{Z}_{[\mu_1 \mu_2]}$
  and $\widetilde{Z}_{[\mu_1 \ldots \mu_5]}$
corresponding to the tensorial central charges occurring in
 (\ref{lutop1})
\par
{\em ii}) The $496=1+165+330$ additional Euclidean tensorial
central charges
 $ \widetilde{Z}$, $\widetilde{Z}_{[\mu_1 \ldots \mu_3]}$
  and $\widetilde{Z}_{[\mu_1 \ldots \mu_4]}$
which describe the maximal complex Hermitean extension of the set
of real bosonic charges occurring in the Minkowski case. We shall
show, however, that one can find
 a holomorphic subalgebra of (\ref{lutop4})
defining holomorphic Euclidean $M$-algebra with 32 complex supercharges and
 $528$ bosonic generators.

In our paper, in order to be  more transparent, we study at first
in Section $2$ the lower-dimensional case of $D=3$ Euclidean
superalgebra.

It appears that to $528$ bosonic charges of Minkowski $M$-algebra
 (\ref{lutop1}) correspond in $D=3$ just   three bosonic charges
 describing $D=3$ three-momenta (there are no Minkowski central
  charges in $D=3$), and to the extension in
  $D=11$ from $528$ Minkowskian to $1024$ Euclidean bosonic charges corresponds in $D=3$
  the extension of  three
   momentum generators by one additional central charge. We see therefore
    that, when passing from
    $D=3$ to $D=11$, instead of
  one $D=3$ Euclidean central charge we obtain $496$ central charges
  in $D=11$.

In Section $3$ we study in more detail  the $D=11$ case and
 in particular the $D=11$ tensorial structure of $496$ Euclidean
  central charges. We introduce in $D=3$ and $D=11$
   the  Osterwalder-Schrader conjugation which is required
    if we wish to obtain the  holomorphic Euclidean $M$-theory with real
     bosonic charges.
   In Section $4$ we present an outlook,
considering in particular the possible applications
 to $D=11$ Euclidean superbrane scan
   as well as  the Euclidean version of the
generalized AdS, dS and conformal superalgebras.
 We would also like to recall here that recently Euclidean
 symmetry and Euclidean superspace was considered as a basis for
 noncommutative supersymmetric field theory
 \cite{ref18}--\cite{ref20}.

\section{The $D=3$ Euclidean superalgebra and the role of  quaternionic
 and  complex structure.}
\sec

The $D=3$ Euclidean spinors are described by real quaternions
 $(q_0, q_r \in R; r=1,2,3)$
 \begin{equation}\label{lutop2.1}
    q = q_0 + e_r \, q_r \qquad e_r e_s = - \delta_{rs} + \varepsilon_{rst}
    e_t \, ,
\end{equation}
which carry the representations of $\overline{SO(3)} = Sp(1) = U(1;{\bf H})$, described
 by unit quaternions
\begin{equation}\label{lutop2.2}
    \alpha = \alpha_0 + \alpha_r e_r \, , \qquad
    \alpha^2_0 + \alpha^2_r =1 \, .
\end{equation}
The quaternionic spinor (\ref{lutop2.1}) is modified under
$Sp(1)$ transformation law as follows
\begin{equation}\label{lutop2.3}
    q' =\alpha\,q \, .
\end{equation}
One can describe the real quaternion $q \in {\bf H}(1)$ by a pair of complex
variables $(z_1,z_2)$. Further, introducing the $2\times 2$ complex matrix
 representation $e_r = -i\sigma_r$, one can represent unit quaternions
  (\ref{lutop2.2}) as $2\times 2$ unitary matrices $A$:
  \begin{equation}\label{lutop2.4}
    \alpha \Leftrightarrow A = \alpha_0 {\bf 1} - i \alpha_r \, \sigma_r \, ,
\end{equation}
i.e. $Sp(1)\simeq  SU(2)$.

We should assume that $D=3$ Euclidean supercharges are the $SO(3)$
spinors. Unfortunately, since the Clifford algebra
\begin{equation}\label{lutop2.5}
    \{  \widetilde{\gamma}_r, \widetilde{\gamma}_s\} = 2 \delta_{rs} \, ,
    \qquad r,s=1,2,3
\end{equation}
has the fundamental ${\bf C}^2 \times {\bf C}^2$ representation, one can not
 employ single quaternionic supercharges as describing the Hermitean
  $D=3$ $N=1$ Euclidean superalgebra. In fact, if we introduce the quaternionic
   Hermitean superalgebra with supercharges described by
   fundamental $SO(3)$ spinor
   \begin{equation}\label{lutop2.6}
    \{ \overline{\bf R}, {\bf R} \} = {\bf Z} \, ,
\end{equation}
where ${\bf R}=R_0 + e_r R_r \to \overline{\bf R} = R_0 -e_r R_r$
 describes the
quaternionic conjugation, it will contain only \underline{one}
bosonic charge
 ${\bf Z}\in {\bf H}$ $({\bf Z} =\overline{\bf Z}
   \to {\bf Z} = Z_0$) and can be
 successfully used rather for the description of $D=1$ $N=4$ supersymmetric quantum mechanics
 \cite{ref18}. In order to obtain the ``supersymmetric roots" of $D=3$ Euclidean momenta one should
 introduce however,
 in agreement with the representation theory of $D=3$ Euclidean
 Clifford algebra (\ref{lutop2.5}),
 two complex supercharges $(\widetilde{Q}_1, \widetilde{Q}_2)$.
  One can write
 the $D=3$ Euclidean superalgebra
  in the following familiar complex-Hermitean form $(\alpha,\beta=1,2)$
 \begin{eqnarray}\label{lutop2.7}
\left\{ \widetilde{Q}_\alpha^\ast , \widetilde{Q}_\beta \right\} & = &
 (\sigma_r)_{\alpha\beta}\, \widetilde{P}_r + \delta_{\alpha\beta} \, Z \, ,
\cr \cr
 \left\{ \widetilde{Q}_\alpha , \widetilde{Q}_\beta \right\} & = &
\left\{  \widetilde{Q}_\alpha^\ast  , \widetilde{Q}_\beta^\ast \right\} = 0\, .
\end{eqnarray}
We see that
 among the bosonic generators
 besides the three momenta we obtain a fourth real central charge ${ Z}$. In
fact (\ref{lutop2.7}) can be obtained by dimensional reduction from
 standard $D=4$ $N=1$ super-algebra.

 In order to find the  quaternionic structure in the superalgebra
  (\ref{lutop2.7}) one should introduce the following pair of two-component
  spinors $R_\alpha, R_\alpha^H$
 \begin{eqnarray}\label{lutop2.8}
R_\alpha & = & \frac{1}{\sqrt{2}} \left(
\widetilde{Q}_\alpha +  \varepsilon_{\alpha\beta} \widetilde{Q}_\beta^\ast
\right)\, ,
\cr\cr
R_\alpha^H & = & \frac{i}{\sqrt{2}} \left(
\widetilde{Q}_\alpha -  \varepsilon_{\alpha\beta} \widetilde{Q}_\beta^\ast
\right)\, ,
\end{eqnarray}
satisfying the relation
\begin{equation}\label{lutop2.9}
    R^H_\alpha = -i \, \varepsilon_{\alpha\beta} \, R_\beta^\ast \, .
\end{equation}
The formula (\ref{lutop2.9}) implies  quaternionic
 reality condition in complex framework  \cite{ref21bis}--\cite{ref19a}
 described by
  $SU(2)$-Majorana condition.
Indeed, introducing $R^1_\alpha = R_\alpha$,  $R^2_\alpha =
R_\alpha^H$
 one can rewrite (\ref{lutop2.9}) in the following way \cite{ref21bis}
 \begin{equation}\label{lutop2.11bis}
    R^a_\alpha = i\, \varepsilon^{ab} \, \varepsilon_{\alpha\beta}
     (R^b_\beta )^\ast \, .
\end{equation}
 The self-conjugate super-algebra (\ref{lutop2.7}) can be
 written as follows:
\bl
 \begin{eqnarray}\label{lutop2.10a}
\left\{  {R}_\alpha , {R}_\beta \right\}
 & = &
 (\varepsilon \sigma_r)_{\alpha\beta} \, \widetilde{P}_r \, ,
 \\ \cr
 \label{lutop2.10b}
 \left\{  {R}_\alpha , {R}_\beta^H \right\}
 & = &
 i \, \varepsilon_{\alpha\beta} \, Z \, ,
 \\ \cr
 \label{lutop2.10c}
 \left\{  {R}_\alpha^H , {R}_\beta^H \right\}
 & = &
  (\varepsilon\sigma_r)_{\alpha\beta} \, \widetilde{P}_r \, ,
\end{eqnarray}
\el
and describes the $N=(1,1)$ $D=3$ Euclidean supersymmetry.

 It is easy to check that the equations
 (\ref{lutop2.10a}--\ref{lutop2.10c}) are consistent with the
 relation (\ref{lutop2.9}), i.e. the $SU(2)$-Majorana reality
  condition (\ref{lutop2.11bis}) can be imposed.

The real superalgebra (\ref{lutop2.10a}--\ref{lutop2.10c})
contains as subsuperalgebras the holomorphic $N=(1,0)$ superalgebra
(\ref{lutop2.10a}) as well as the antiholomorphic $N=(0,1)$
superalgebra (\ref{lutop2.10c}).
  In particular
  we can consider separately (\ref{lutop2.10a}) as describing the basic
  $N=1$ $D=3$ Euclidean non-self-conjugate superalgebra,
    supersymmetrizing
   three Euclidean momenta  $P_r$. If the superalgebras (\ref{lutop2.10a})
   or (\ref{lutop2.10c}) are considered as independent algebraic
    structures, the Euclidean three-momentum $P_r$ can be complex.

   The superalgebra (\ref{lutop2.10a}) can be obtained by analytic
   continuation of the real $D=3$  $N=1$ Minkowski superalgebra
    $(\mu=0,1,2$;  $Q_\alpha \in R)$
\begin{equation}\label{lutop2.11}
\left\{ {Q}_\alpha , {Q}_\beta \right\}
= (C \gamma_\mu)_{\alpha\beta} \, P^\mu \, ,
\end{equation}
where $\{\gamma_\mu, \gamma_\nu \} = 2\eta_{\mu\nu}$
 ($\eta_{\mu\nu} = diag(-1,1,1)$, $C=\gamma_0$, $\gamma_\mu \in {\bf R}^2 \times
 {\bf R}^2$)
  and one can choose e.g.
  \begin{equation}\label{lutop2.12}
    \gamma_0 = - i \sigma_2 \, , \quad
    \gamma_1 = \sigma_1\, , \quad
    \gamma_2 = \sigma_3 \,.
\end{equation}
 The analytic continuation of the
 Minkowski superalgebra (\ref{lutop2.11}) to the Euclidean one, given
by (\ref{lutop2.10a}), is obtained
 by complexification of the real $D=3$ Minkowski spinors and Wick rotation
  of Minkowski vectors into Euclidean ones.
We
 replace ${Q}_\alpha \in {\bf R}$ by $\widetilde{Q}_\alpha  \in {\bf C}$ and
 \begin{eqnarray}\label{lutop2.13}
   & P^0 \Rightarrow \widetilde{P}_3 = i\, P^0 \, ,
    \qquad
    &\widetilde{P}_k = P_k \qquad \qquad  \ (k=1,2) \, ,
\cr
 & \gamma^0 \Rightarrow \widetilde{\gamma}_3= i \, \gamma_0=
\sigma_2 \qquad & \widetilde{\gamma}_k = \gamma_k =\sigma_k \qquad
(k=1,2)
\end{eqnarray}
where $\widetilde{C}= 
 \varepsilon = i \gamma_2$  
and $\widetilde{C} \widetilde{\gamma}_r =
- \widetilde{\gamma}_r^T \widetilde{C}$,
 i.e. we keep the same  $D=3$ charge conjugation matrices in
  Euclidean and Minkowski case.
  One gets $(r=1,2,3)$
 \begin{equation}\label{lutop2.15}
    \left\{ R_\alpha, R_\beta \right\} = \left(
\widetilde{C}\, \widetilde{\gamma}_r
    \right)_{\alpha\beta}\, \widetilde{P}_r
\end{equation}
i.e.  one can identify after putting $Q^E_\alpha = R_\alpha$ the
superalgebras (\ref{lutop2.15}) and (\ref{lutop2.10a}).

In order to justify the real values of $P_r$ in (\ref{lutop2.15})
 one should introduce the Osterwalder-Schrader $(OS)$ conjugation
  $A \to A^{\#}$, which is defined in any dimension $D$
  with complex Euclidean spinors replacing real Majorana spinors
    as complex
   conjugation   supplemented by
     time reversal transformation, i.e.
   \begin{equation}\label{lutop2.16bis}
    R^{\#}_A = T_{AB}\, R^\ast_B
\end{equation}
where ($D$ is the Euclidean time direction; following
(\ref{lutop2.13}) we choose $\widetilde{k}_{\mu}$ $(k=1,
\ldots ,D-1)$ real and $\Gamma_D$ purely imaginary)
\begin{eqnarray}\label{lutop2.17bis}
    T \, \widetilde{\Gamma}_k \, T^{-1} =  \widetilde{\Gamma}_k \, ,
     \qquad
     T  \widetilde{\Gamma}_D  {T}^{-1} =  \widetilde{\Gamma}_D\, .
\end{eqnarray}
The  relation (\ref{lutop2.17bis}) implies that
\begin{eqnarray}\label{lutop2.18bis}
\widetilde{P}^{\#}_k =
 \widetilde{P}_k \, ,
\qquad
 \widetilde{P}^{\#}_D  =
 - \widetilde{P}_D \, .
\end{eqnarray}
It appears that in $D=3$ the $OS$ conjugation of supercharges can
 be identified with the conjugation
 (\ref{lutop2.9}), i.e. $T_{\alpha\beta} = -i \varepsilon_{\alpha\beta}$,
 $\widetilde{\Gamma}_{k} = (\sigma_1,\sigma_3)$ and $\widetilde{\Gamma}_D =\sigma_2$.

  The real values of ${\widetilde P}_r$ in (\ref{lutop2.15}) are required if we assume the invariance
   of the superalgebra (\ref{lutop2.15}) under $OS$ conjugation,
    i.e. the conjugation (\ref{lutop2.9}). This reality
    requirement is satisfied inside the superalgebra
    (\ref{lutop2.10a}--\ref{lutop2.10c}) and is
     equivalent to the consistency of (\ref{lutop2.10a}--\ref{lutop2.10c})
       with subsidiary condition (\ref{lutop2.11bis}).
    Indeed
    performing such a conjugation one obtains from
     the $N=(1,0)$ superalgebra (\ref{lutop2.10a})
   the identical in its form  $N=(0,1)$ superalgebra (\ref{lutop2.10c}).

One can therefore say that

{\em i}) our $D=3$ $ N=(0,1)$  Euclidean superalgebra
 (\ref{lutop2.10c}) is $OS$-conjugate to the superalgebra
(\ref{lutop2.10a}) (or equivalently (\ref{lutop2.15})).

{\em ii}) It is not possible to impose the $OS$ reality conditions
on
 single pair of  supercharges $R_\alpha$
 \begin{equation}\label{lutop2.17}
    \left( R_\alpha \right)^{\#}
     = R_\alpha \quad \Longleftrightarrow \quad
     R_\alpha = i\,  \varepsilon_{\alpha\beta}
     \left( R_\alpha \right)^\ast \, ,
\end{equation}
because such a condition is not consistent, i.e. the superalgebra
 $(1,0)$ can not be made selfconjugate.

\section{The Euclidean $M$-algebra and the role of quaternionic
 and  complex structure.}
\sec

In this Section we shall translate the Euclidean superalgebra
  structures from $D=3$ to
$D=11$. Due to the Bott's periodicity conditions these algebraic structures
 should be analogous.

 The $D=11$ Euclidean spinors are described by $16$ quaternions
  $R_m \in {\bf H}^{16}$ ($m=1, \ldots, 16)$ and the fundamental reprsentation
  of the $D=11$ Euclidean Clifford algebra (\ref{lutop3}) is
   ${\bf C}^{32} \times {\bf C}^{32} $.
     The quaternionic $D=11$ Hermitean superalgebra,
    generalizing relation (\ref{lutop2.6}), is given by the relation
    \begin{equation}\label{lutop3.1}
    \left\{ \overline{\bf R}_m, {\bf R}_n \right\} = {\bf Z}_{mn}\, ,
    \qquad
    {\bf Z}_{mn}= \overline{\bf Z}_{nm} \, .
\end{equation}
The $16\times 16$ Hermitean-quaternionic matrix $Z_{mn}$ is described
 by $496$ real bosonic Abelian generators.

  In order to describe the complex-Hermitean Euclidean $M$-algebra we should
   introduce $32$ complex supercharges $\widetilde{Q}_A \in {\bf C}^{32}$.
   The most general complex-Hermitean $D=11$ Euclidean algebra is given
    by the relation
    \begin{equation}\label{lutop3.2}
    \left\{ \widetilde{Q}_A^+, \widetilde{Q}_B \right\} = \widetilde{Z}_{AB}\, ,
    \qquad
   \widetilde{Z}_{AB}= \widetilde{Z}_{AB}^+ \, ,
\end{equation}
containing $1024$ real bosonic charges.
We introduce $D=11$ Euclidean gamma-matrices by putting
\begin{eqnarray}\label{lutop3.3}
    \widetilde{\Gamma}_\mu & =  & \Gamma _\mu \, \qquad
    \mu=1,2,\ldots 10
    \cr
    \widetilde{\Gamma}_{11} & =  & i \Gamma_0
\end{eqnarray}
where the matrices $\Gamma _\mu, \Gamma _0$ describe the real $32$-dimensional
 Majorana representation of the $D=11$ Minkowskian Clifford algebra
 (\ref{lutop2}) occurring in the  $M$-algebra (\ref{lutop1}).
 If we introduce the antisymmetric products
$ \widetilde{\Gamma}_{[\mu_1 \ldots \mu_p]} =
 \frac{1}{p!}\sum\limits_{(\mu_1 \ldots \mu_p)}
 (-1)^{perm}
 \widetilde{\Gamma}_{\mu_1} \ldots \widetilde{\Gamma}_{\mu p}$ we can
 check that
 \begin{eqnarray}\label{lutop3.4}
 \widetilde{\Gamma}_{[\mu_1 \ldots \mu_p]}^+
  & = &
\widetilde{\Gamma}_{[\mu_1 \ldots \mu_p]} \qquad
 \hbox{for} \quad p=0,1,4,5,8,9
 \cr
\widetilde{\Gamma}_{[\mu_1 \ldots \mu_p]}^+
  & = &
- \widetilde{\Gamma}_{[\mu_1 \ldots \mu_p]} \qquad
 \hbox{for} \quad p=2,3,6,7,10,11 \, .
\end{eqnarray}
Following the symmetry properties (\ref{lutop3.4}) one
 can present  the superalgebra
 (\ref{lutop3.2}) in the form (\ref{lutop4}) or (\ref{lutop1.7}).

 In order to write down the Euclidean $M$-algebra with only $528$ bosonic
  charges - the ones
 corresponding to the bosonic charges
   of the $M$-algebra
    (\ref{lutop1}) - one should exhibit in $D=11$ the quaternionic
    or $SU(2)$-Majorana structure in analogous way as in the eq.
     (\ref{lutop2.10a}--\ref{lutop2.10c}) for the $D=3$ case.

     Let us introduce the following pair of $32$-component complex
     supercharges:
     \begin{eqnarray}\label{lutop3.5}
     R_A & = & \frac{1}{\sqrt{2}} \left(
      \widetilde{Q}_A + C_{AB} \, \widetilde{Q}_B^\ast
      \right)
      \cr
      R_A^H & = & \frac{i}{\sqrt{2}} \left(
      \widetilde{Q}_A - C_{AB} \, \widetilde{Q}_B^\ast
      \right)
\end{eqnarray}
where the $D=11$   charge conjugation matrix satisfies the relations
\begin{equation}\label{lutop3.6}
    C\widetilde{\Gamma}_\mu = - \widetilde{\Gamma}_\mu^T \, C \,
     \qquad
     C^2 = - 1 \, \qquad C^T = - C\, ,
\end{equation}
and can be chosen the same in Minkowski and Euclidean case.

One can show that
\begin{equation}\label{lutop3.7}
    R^H_A = -i \, C_{AB} \, R^\ast_B \,.
\end{equation}
Introducing $64$-component complex spinor
$R^a_A =(R_A, R^H_A)$ one can rewrite
(\ref{lutop3.7}) as the $SU(2)$-Majorana condition \cite{ref13}
\begin{equation}\label{lutop3.8}
R^a_A = i \varepsilon^{ab} \, C _{AB} (R^b_B)^\ast \,.
\end{equation}
The superalgebra (\ref{lutop1.7}) can be written in the following form
$(C=\Gamma_0$; $\mu=1,\ldots 11)$
\bl
\begin{eqnarray}\label{lutop3.9a}
\left\{ R_A, R_B\right\}  & = &
 \left( C\widetilde{\Gamma}_\mu\right)_{AB}
 \widetilde{P}_\mu
 + \left(
 C\widetilde{\Gamma}_{[\mu\nu]}\right)_{AB}
 \widetilde{Z}_{[\mu\nu]}
 \cr
 && +  \left( C\widetilde{\Gamma}_{[\mu_1 \ldots \mu_5]}\right)
 \widetilde{Z}_{[\mu_1 \ldots \mu_5]}
 \\
 \label{lutop3.9b}
\left\{ R_A, R_B^H\right\}  & = & i
\left\{
 C_{AB} \widetilde{Z} + \left(
C\widetilde{\Gamma}_{[\mu_1\mu_2\mu_3]} \right)_{AB}
 \widetilde{Z}_{[\mu_1\mu_2\mu_3]}
  \right.
 \cr
 &&
\left.
 \left( C\widetilde{\Gamma}_{[\mu_1 \ldots \mu_4]}
 \right)_{AB}
 \widetilde{Z}_{[\mu_1 \ldots \mu_4]}
\right\}
 \\
 \label{lutop3.9c}
\left\{ R_A^H, R_B^H\right\}  & = &
 \left( C\widetilde{\Gamma}_{\mu}\right)_{AB}
 \widetilde{P}_\mu
  +
 \left( C\widetilde{\Gamma}_{[\mu\nu]}\right)_{AB}
\widetilde{Z}_{\mu\nu} +
\left(
 C \widetilde{\Gamma}_{[\mu_1 \ldots \mu_5]}
 \right)_{AB}
 \widetilde{Z}_{[\mu_1 \ldots \mu_5]}\, .
\end{eqnarray}
\el
\indent
We see that

{\em i}) The relation (\ref{lutop3.9a}--\ref{lutop3.9c}) describe the
selfconjugate $(1,1)$ Euclidean $M$-algebra with $1024$ bosonic
 Abelian charges which can be written also in the form
(\ref{lutop4})

{\em  ii}) The relation (\ref{lutop3.9a}) describes the holomorphic
$(1,0)$ Euclidean $M$-algebra, with $528$ Abelian bosonic charges.
The antiholomorphic $(0,1)$ Euclidean $M$-algebra obtained by
 conjugation (\ref{lutop3.7}), which contains also
  $528$ Abelian bosonic charges, is given by the relation
  (\ref{lutop3.9c}).

{\em iii}) The subsidiary condition (\ref{lutop3.8}) relates the bosonic
 generators of the $(1,0)$ and
  $(0,1)$ sectors of the superalgebra
  (\ref{lutop3.9a}--\ref{lutop3.9c}). It should be observed that
  the $496$ real
    bosonic generators occurring in quaternionic superalgebra
   (\ref{lutop3.1}) can
     be found in the ``cross anticommutator" (\ref{lutop3.9b}).

   By analogy with the $D=3$ case one can treat the $(1,0)$ holomorphic
   Euclidean $M$-algebra
   (\ref{lutop3.9a}) as analytic continuation of the $D=11$ Minkowski
    $M$-algebra, given by
    (\ref{lutop1}). For such purpose one should in
    (\ref{lutop1}) complexify  the supercharges and perform the Wick
     rotation of Minkowski $11$-tensors, i.e. perform the
    change $\Gamma_\mu \to \widetilde{\Gamma}_\mu$ (see
      (\ref{lutop3.3})) supplemented by the redefinitions
       $(k,l\ldots = 1,2, \ldots ,10 )$:
 \begin{eqnarray}\label{lutop3.11}
    \widetilde{P}_k  =  P_k
    \qquad
    \widetilde{Z}_{[k\, l]} = {Z}_{[k\, l]}
    \qquad
\widetilde{Z}_{[k_1 \ldots k_5]} =
{Z}_{[k_1 \ldots k_5]}
\cr
\widetilde{P}_{11}  =  i P_0
\qquad
    \widetilde{Z}_{[k {\, 11}]} = i {Z}_{[k0]}
    \qquad
\widetilde{Z}_{[11 k_1 \ldots k_4]} =
 i \, {Z}_{[0 k_1 \ldots k_4]}\, ,
\end{eqnarray}
changing Minkowski tensors into Euclidean tensors.

In such a way we shall obtain from  (\ref{lutop1}) the holomorphic
Euclidean $M$-algebra (\ref{lutop3.9a}).

The quaternionic conjugation (\ref{lutop3.7}) describes the $D=11$
 $OS$ conjugation in Euclidean space.
  We would like to point out that in
holomorphic and antiholomorphic
 Euclidean $M$-algebra (i.e. if we consider
 the relations (\ref{lutop3.9a}) and (\ref{lutop3.9c}) as separate)
the bosonic generators, in particular the $11$-momenta, can be
complex.

In order to obtain e.g. in (\ref{lutop3.9a}) the real
Abelian bosonic generators
 \begin{eqnarray*}\label{lutopempty}
\widetilde{P}_k = P_k^\ast  \, , \qquad
 \widetilde{Z}_{[k\, l]} = Z^\ast_{[k\, l]}\, ,
 \qquad
\widetilde{Z}_{[k_1 \ldots k_5]}
  = Z^\ast_{[k_1 \ldots k_5]}\, .
\end{eqnarray*}
 one should impose the invariance of
 the superalgebra  (\ref{lutop3.9a}) under the $OS$ conjugation (\ref{lutop3.7}),
   i.e.  assume that the form of holomorphic and antiholomorphic Euclidean
    $M$-algebras related by $OS$ conjugation is identical.

\section{Conclusions.}
\sec

In relation with Euclidean $M$-theories and their algebraic
description presented in this paper we would like to make the
following comments:

{\em i}) It is well-known \cite{ref9} that the presence of tensorial central
charges in generalized $D$-dimensional supersymmetry algebra can
be linked with the presence of $p$-brane solution of
$D$-dimensional supergravity. We have two $D=11$ Euclidean
$M$-theories described by (1,1) self-dual (see (\ref{lutop2.7}))
and $(1,0)$ holomorphic (see (\ref{lutop2.10a}) or
(\ref{lutop2.15})) supersymmetry algebras.
In Euclidean theory the role of $p$-brane solutions will be played by Euclidean
 instantons and space branes ($S$-branes).
 It appears that in holomorphic $N=(1,0)$ Euclidean $M$-theory the set of
 instanton solutions  corresponds to $p$-dimensional solutions in
   standard Minkowskian $D=11$
$M$-theory ($2$-branes and $5$-branes, supplemented by six-dimensional Kaluza-Klein monopoles
and nine-dimensional Ho\v{r}ava-Witten boundaries). The Euclidean $N=(1,1)$ $M$-theory with
symmetry algebra (\ref{lutop2.7})
 will have additional instanton or $S$-brane
   solutions corresponding to $3$-tensor
  and $4$-tensor central charges
  which do not have  their Minkowski  space counterparts.

{\em ii}) The superalgebras either with Hermitean selfconjugate algebra
structure or holomorphic structure can be considered in any
dimension with complex fundamental spinors ($D=0,4$ modulo $8$
for Minkowski metric and $D=2,6$ modulo $8$ for Euclidean metric)
or fundamental quaternionic spinors ($D=5,6,7$ modulo $8$
for Minkowski metric and $D= 3,4,5$ modulo $8$ for Euclidean
metric). The physical choice of the algebraic structure  of
 supersymmetry is
indicated by the presence  in the bosonic sector of the vectorial
momentum generators. For example in $D=4$ one can choose either the
Hermitean algebra ($\alpha,\beta=1,2$)
\begin{equation}\label{lutop4.1}
    \left\{ Q^+_{\dot{\alpha}}, Q_\beta \right\}
     = P_{\dot{\alpha} \beta}\, ,
\end{equation}
or the pair of holomorphic$/$antiholomorphic algebras:
\begin{eqnarray}\label{lutop4.2}
\left\{ Q_{{\alpha}}, Q_\beta \right\}
    & = &
     Z_{\alpha {\beta}}\, ,
     \cr
     \left\{ Q^+_{\dot{\alpha}}, Q^+_{\dot{\beta}} \right\}
    & = &   Z_{\dot{\alpha} \dot{\beta}}
    = Z^+_{\alpha {\beta}}\, ,
    \cr
     \left\{ Q^+_{\dot{\alpha}}, Q_\beta \right\}
     & = & 0 \, .
\end{eqnarray}
Since the four-momentum generators are present only in the relation
(\ref{lutop4.1}), these relations are the basic $D=4$ supersymmetry
relations.

In the quaternionic $D=3$ and $D=11$ Euclidean case the Hermitean
 algebras  (\ref{lutop2.10b}) and (\ref{lutop3.9b})
 do not contain the momentum
generators; these generators occur in the superalgebras
(\ref{lutop2.10a},\ref{lutop2.10c}) and (\ref{lutop3.9a},\ref{lutop3.9c}).
  We see therefore that in these cases the holomorphic/antiholomorphic
   algebra is more physical.

{\em iii}) If the fundamental spinors are complex, from algebraic
 point of view one can consider the
minimally extended supersymmetry algebra with either Hermitean or
holomorphic complex structure. If we assume however that
the Hermitean anticommutator $\{Q^+_A,Q_B\}$ as well as
the holomorphic one $\{Q_A,Q_B\}$ are  saturated by Abelian bosonic
generators (tensorial central charges), we obtain the most general
real superalgebra. In such a way in $D=4$ Minkowski case
 six tensorial
central charges are generated by the relations (\ref{lutop4.2}), while the
 Hermitean superalgebra (\ref{lutop4.1})
describes only the four-momentum generators.

 In quaternionic case we have two levels of
generalizations. Assuming that the fundamental spinors belong to
 ${\bf H}^n$, one can consider:

 1) generalized Hermitean superalgebra for the supercharges
 belonging to ${\bf C}^{2n}$. In such a way we generate $4n^2$ real
 Abelian bosonic generators.

 2) One can write down also the most general real superalgebra, with
 supercharges belonging to ${\bf R}^{4n}$. In such a way we obtain
 $2n(4n+1)$ real Abelian generators.

 In $D=11$ Euclidean case $n=16$ the generalized complex-Hermitean
 superalgebra is given by the relation
 (\ref{lutop2.7}); the minimal real  superalgebra containing
  (\ref{lutop2.7}) takes the form
  $(S,T=1, \ldots ,
 64)$
 \begin{eqnarray*}\label{lutopnic}
 \left\{ Q_S, Q_T \right\} = P_{ST} \, ,
\end{eqnarray*}
and can be obtained by the contraction of $OSp(1|64;{\bf R})$.

{\em iv}) In this paper we discussed the analytic continuation of $D=11$
Minkowski superalgebra with only supersymmetrized  Abelian bosonic
charges. It has been argued (see e.g. \cite{ref24}-\cite{ref26}), due to the relation
$Sp(32)\supset O(10,2)$, that $OSp(1|32;{\bf R})$ is the generalized
$D=11$ $AdS$ superalgebra, and $OSp(1|64;{\bf R})$ describes the generalized
$D=11$ superconformal algebra. If we consider the holomorphic
version of Euclidean $M$-theory the corresponding generalized
Euclidean $AdS$ superalgebra and generalized Euclidean conformal
superalgebra are
obtained by holomorphic complexification and
respectively given by the $OSp(1|32;{\bf C})$ and
$OSp(1|64;{\bf C})$ superalgebras.

{\em v}) It appears that the $M$-algebra (\ref{lutop1}) describes as
well the symmetries of nonstandard $M^*$-theory with signature
(9,2), and $M'$-theory, with signature (6,5), which are related
with standard $M$-theory by dualities (\cite{ref21,ref22}; see also
 \cite{ref23}).
Different choices of
signature corresponds to different choices of
 holonomy groups embedded in
$sl(32;{\bf R})$. If we pass to complex Hermitean holonomy structures,
embedded in $sl(32;{\bf C})$, we can describe $12$ different versions of
$M$-theories, with any signature ($11-k,k$) ($k=0,1\ldots 11$).
In particular, for signatures (8,3) and (4,7) we should use the
holonomy groups embedded in $Sl(16;{\bf H})\subset Sl(32;{\bf C})$.

\subsection*{Acknowledgments}

The authors would like to thank D. Sorokin for valuable remarks.


\begin{thebibliography}{99}
\bibitem{ref1} R. Cameron. J. Anal. Math. {\bf 10}, 287, (1962/63).

\bibitem{ref2} B. Simon, "The $P(\Phi)_2$ (Euclidean) Quantum
Field Theory", Princeton Univ. Press. 1974.

\bibitem{ref3} G.W. Gibbons, S.W. Hawking and M.J. Perry, Nucl. Phys.
{\bf B138}, 141 (1978).

\bibitem{ref4} A.V. Belitsky, S. Vandoren and P. van Nieuwenhuizen,
 Phys. Lett. {\bf 477}, 335 (2000).

\bibitem{ref5} G. Parisi and Wu, Scientica Sinica {\bf 24}, 483 (1981).

\bibitem{ref6} E. Nelson, ``Quantum Fluctuations'', Princeton Univ. Press,
 New Jersey, 1985.

\bibitem{ref7} M. J. Duff, Int. J. Mod. Phys. {\bf A11}, 5623 (1996).

\bibitem{ref8} E. Witten, J. Geom. Phys. {\bf 22}, 1 (1997); {\bf 22}, 103
 (1997).

\bibitem{ref9}J.A. de Azcarraga, J.P. Gauntlett, J.M., Izquierdo
and P.K. Townsend, Phys. Rev. Lett. {\bf 63}, 2443 (1989).

\bibitem{ref10} P. Townsend, Cargese Lectures, 1997;
hep-th/9712004.

\bibitem{ref11} R. G\"{u}ven, Phys. Lett. {\bf B276}, 49 (1992).


\bibitem{ref12} K.S. Stelle, ``Lectures on Supergravity p-Branes'',
 hep-th/9701088.


\bibitem{ref13}B. Zumino, Phys. Lett. {\bf B69}, 369 (1977).

\bibitem{ref14} J. Lukierski, A. Nowicki, Journ. Math. Phys. {\bf
25}, 2545 (1984).

\bibitem{ref15}J. Lukierski, Czech. Journ. Phys. {\bf B37}, 359 (1987).

\bibitem{ref16} K. Osterwalder and R. Schrader, Helv. Phys.
 Acta {\bf 46}, 277 (1973).

\bibitem{ref17} J. Fr\"{o}hlich and K. Osterwalder, Helv. Phys.
 Acta {\bf 47}, 781) (1974).

\bibitem{ref18} N. Seiberg, JHEP 0307 (2003) 045.


 \bibitem{ref20a} E. Ivanov, O. Lechtenfeld and B. Zupnik,
 hep-th/00308012.

 \bibitem{ref20} S. Ferrara and E. Sokatchev, hep-th/0308021.

\bibitem{ref21bis} T. Kugo, P. Townsend, Nucl. Phys. {\bf B221},
357 (1983).

\bibitem{ref19} J. Lukierski and A. Nowicki, Ann. Phys. {\bf 166},
 164 (1986).

\bibitem{ref19a} F. Toppan, Nucl. Phys. (Proc. Suppl.) {\bf B 102-103}, 270 (2001).


\bibitem{ref24} J.W. van Holten and A. van Proyen, J. Phys. {\bf A15},
 3763 (1982).

 \bibitem{ref25} I. Bars, Phys. Lett. {\bf B457}, 275 (1999);
 ibid. {\bf B483}, 248 (2000).

 \bibitem{ref26} I. Bandos, J. Lukierski, D. Sorokin,
  hep-th/9912051.

 \bibitem{ref21} C.M. Hull, J. High Energy Phys. {\bf 9811}, 017 (1999).

 \bibitem{ref22} C.M. Hull and R.R. Khuri, Nucl. Phys. {\bf B536}, 219 (1998);
  {\bf B575}, 231 (2000).

 \bibitem{ref23} M.A. De Andrade, M. Rojas and F. Toppan, Int. J. Mod.
  Phys. {\bf A16}, 4453 (2001).


\end{thebibliography}
\end{document}